\documentclass[a4paper,10pt]{article}
\usepackage{graphicx}
\usepackage{comment}
\usepackage{amssymb}
\usepackage{amsmath}
\usepackage{nicefrac}
\usepackage{graphicx}
\usepackage{dcolumn}
\usepackage{bm}
\usepackage{comment}
\usepackage{multirow}
\usepackage{cite}
\usepackage{mathrsfs}

\begin{document}

\newcommand{\threej}[6]{\left(\begin{array}{ccc}#1 & #2 & #3 \\ #4 & #5 & #6 \end{array}\right)}
\newcommand{\six}[6]{\left\{\begin{array}{ccc}#1 & #2 & #3 \\ #4 & #5 & #6 \end{array}\right\}}

\huge

\begin{center}
Inequalities for exchange Slater integrals
\end{center}

\vspace{0.5cm}

\large

\begin{center}
Jean-Christophe Pain$^{a,b,}$\footnote{jean-christophe.pain@cea.fr}
\end{center}

\normalsize

\begin{center}
\it $^a$CEA, DAM, DIF, F-91297 Arpajon, France\\
\it $^b$Universit\'e Paris-Saclay, CEA, Laboratoire Mati\`ere sous Conditions Extr\^emes,\\
\it 91680 Bruy\`eres-le-Ch\^atel, France
\end{center}

\vspace{0.5cm}

\begin{abstract}
The variations of exchange Slater integrals with respect to their order $k$ are not well known. While direct Slater integrals $F^k$ are positive and decreasing when the order increases, this is not {\it stricto sensu} the case for exchange integrals $G^k$. However, two inequalities were published by Racah in his seminal article ``Theory of complex spectra. II''. In this article, we show that the technique used by Racah can be generalized, albeit with cumbersome calculations, to derive further relations, and provide two of them, involving respectively three and four exchange integrals. Such relations can prove useful to detect regularities in complex atomic spectra and classify energy levels. 
\end{abstract}

%\maketitle

\section{Introduction}\label{sec1}

In order calculate the energy of electronic configurations in atomic physics and spectroscopy \cite{Sobelman1992}, one has to determine several matrix elements, representing the kinetic energy of the electrons, the electron-nucleus interaction, the electron-electron interaction, the spin-orbit interaction, or the spin-spin interaction among the most important. The matrix element of the Coulomb electrostatic operator reads \cite{Cowan1981}:
\begin{eqnarray}\label{svd}
\langle ij\vert\frac{1}{r_{12}}\vert tu\rangle
&=&\delta_{m_{s_i},m_{s_t}}\delta_{m_{s_j},m_{s_u}}\sum_{k=0}^{\infty}R^k(ij,tu)\sum_{q=-k}^k\delta_{q,m_{\ell_t}-m_{\ell_i}}\delta_{q,m_{\ell_j}-m_{\ell_u}}\nonumber\\
& &\;\;\;\;\times(-1)^qc^k(\ell_i,m_{\ell_i},\ell_t,m_{\ell_t})c^k(\ell_j,m_{\ell_j},\ell_u,m_{\ell_u}), 
\end{eqnarray}
where $r_{12}$ is the distance between electrons 1 and 2 and $\delta$ the Kronecker symbol. The quantity $R^k$ is
\begin{eqnarray}
R^k(ij,tu)&\equiv&\int_0^{\infty}\int_0^{\infty}\frac{r_<^k}{r_>^{k+1}}P_i^*(r_1)P_j^*(r_2)P_t(r_1)P_u(r_2)dr_1 dr_2\nonumber\\
&=&\int_0^{\infty}\left\{\frac{1}{r_2^{k+1}}\int_0^{r_2}r_1^kP_i^*(r_1)P_t(r_1)dr_1\right.\nonumber\\
& &\left.+r_2^k\int_{r_2}^{\infty}\frac{1}{r_1^{k+1}}P_i^*(r_1)P_t(r_1)dr_1\right\} P_j^*(r_2)P_u(r_2)dr_2,
\end{eqnarray}
where $P_i(r)$ is the radial part of the wavefunction multiplied by $r$, and $i$ represents the pair ($n,\ell$), $n$ being the principal quantum number $\ell$ and the orbital quantum number. One has therefore $P_{n\ell}(r)=rR_{n\ell}(r)$. $r_<$ is equal to $\min(r_1,r_2)$ and $r_>$ to $\max(r_1,r_2)$. The matrix element (\ref{svd}) is zero unless $q=m_{\ell_t}-m_{\ell_i}=m_{\ell_j}-m_{\ell_u}$. Mathematically, this means that there is at most one non-zero term in the summation over $q$. Rewriting the latter expression in the form
\begin{equation}
m_{\ell_i}+m_{\ell_j}=m_{\ell_t}+m_{\ell_u}
\end{equation} 
shows that that it represents the conservation of angular momentum: the Coulomb interaction between two electrons can not change the total angular momentum of the electron, not its projections. For the spins one has $m_{s_i}=m_{s_t}$ and $m_{s_j}=m_{s_u}$ (since the electrostatic interaction does not operate on electron spins, not only the total spin angular momentum is conserved, but also the spin of each electron separately). For the direct contribution to the electron-electron matrix element, one has
\begin{equation}
\langle ij\vert\frac{1}{r_{12}}\vert ij\rangle=\sum_{k=0}^{\infty}F^k(ij)c^k(\ell_i m_{\ell_i},\ell_i m_{\ell_i})c^k(\ell_j m_{\ell_j},\ell_j m_{\ell_j})
\end{equation}
where
\begin{equation}\label{fk}
F^k(ij)=R^k(ij,ij)=\int_0^{\infty}\int_0^{\infty}\frac{r_<^k}{r_>^{k+1}}\vert P_i(r_1)\vert^2\vert P_j(r_2)\vert^2dr_1dr_2
\end{equation}
and the exchange contribution simplifies to
\begin{equation}
-\langle ij\vert\frac{1}{r_{12}}\vert ji\rangle=-\delta_{m_{s_i}m_{s_j}}\sum_{k=0}^{\infty}G^k(ij)\left[c^k(\ell_i m_{\ell_i},\ell_j m_{\ell_j})\right]^2,
\end{equation}
where
\begin{equation}\label{gk}
G^k(ij)=R^k(ij,ji)=\int_0^{\infty}\int_0^{\infty}\frac{r_<^k}{r_>^{k+1}} P_i^*(r_1)P_j^*(r_2)P_j(r_1)P_i(r_2)dr_1dr_2.
\end{equation}
The coefficients $c^k$ (angular part) are defined {\it via}

\begin{eqnarray}
& &\left(\frac{4\pi}{2k+1}\right)^{1/2}\int_0^{2\pi}\int_0^{\pi}Y_{\ell m}^*(\theta,\phi)Y_{kq}(\theta,\phi)Y_{\ell' m'}(\theta,\phi)\sin\theta d\theta d\phi\nonumber\\
&&\;\;\;\;\;\;\;\;\;=(-1)^{-m}\left[\ell,\ell'\right]^{1/2}\threej{\ell}{k}{\ell'}{0}{0}{0}\threej{\ell}{k}{\ell'}{-m}{q}{m'}\nonumber\\
&&\;\;\;\;\;\;\;\;\;=\delta_{q,m-m'}~c^k(\ell m,\ell' m'),
\end{eqnarray}
where $Y_{\ell m}(\theta,\phi)$ are the spherical harmonics and $\threej{\ell_1}{\ell_2}{\ell_3}{m_1}{m_2}{m_3}$ the usual $3jm$ symbol. Such integrals involving three spherical harmonics were introduced by Gaunt in 1929 in his study of the triplets of helium, and are usually named after him \cite{Gaunt1929}. In the case where a large number of such integrals must be evaluated (for instance in multiple scattering problems), efficient algorithms are required \cite{Xu1996}.

The radial integrals $F^k$ and $G^k$ (or more generally $R^k$) are referred to as Slater integrals \cite{Slater1929}. They can be evaluated analytically in the case of a Coulomb potential (see for instance Refs. \cite{Naqvi1964,Ruano2013,Hey2017}). Note that to avoid the occurrence of fractional coefficients for the $F_k$ and $G_k$, Condon and Shortley defined $F_k=F^k/\mathscr{D}_k$ and $G_k=G^k/\mathscr{D}_k$ \cite{Condon1935}, where $\mathscr{D}_k$ is the denominator in $c^k(\ell m,\ell' m')=\pm\sqrt{x/\mathscr{D}_k}$ given in their tables 1$^6$ (for $\ell+\ell'$ odd) and 2$^6$ (for $\ell+\ell'$ even), pp. 178 and 179 respectively. 

Besides their importance for the evaluation of the energies of atomic levels and configurations, the exchange Slater integrals (and in particular $G^1$) are known to be responsible for specific properties. For instance, in hot plasma atomic spectra, one can observe a concentration of the oscillator strength towards the high-energy side of the transition array \cite{OSullivan1999,Bauche-Arnoult2000}, which occurs as well for some complex Auger spectra. As an example, for configurations $\ell^N\ell'^{N'+1}$ made of two open subshells having the same principal quantum number, the Coulomb exchange interaction energy plays a major role in the energy level spectrum. The latter interaction is responsible for the formation of the upper and lower groups of levels with very different relative contributions to electronic transitions. Because of the relation between the energy of a level and the transition amplitude of a line involving that level, the most intense lines in the radiative or Auger spectra mainly originate from the upper group of levels. The asymmetrical shape of the transition array comes in general along with a dominant exchange Slater integral $G^1$ with a positive multiplicative coefficient, which is always the case in $\ell^{N+1}\rightarrow \ell^N\ell'$ arrays that are subsequently always asymmetrical. Such a property is strongly connected to the concept of ``emissive zones'' \cite{Bauche-Arnoult1983}. The total strength of all electric-dipole the lines linking an $\alpha J$ level of configuration $\ell^N\ell'^{N'+1}$ to the configuration $\ell^{N+1}\ell'^{N'}$ reads \cite{Bauche2015}:
\begin{eqnarray}
& &S_{E1}\left[\left(\ell^N\ell'^{N'+1}\right)\alpha J-\ell^{N+1}\ell'^{N'}\right]\nonumber\\
& &\;\;\;\;\;\;\;\;\;\;\;\;=(2J+1)\left[\frac{(N'+1)\ell_>}{2\ell'+1}+\mathscr{C}(G^1;\alpha J)\right]I^2(n\ell,n'\ell'),
\end{eqnarray}
where $\mathscr{C}(G^1;\alpha J)$ is the coefficient of the $G^1$ Slater integral in the energy of the $\alpha J$ level, whatever the coupling and
\begin{equation}
I(n\ell,n'\ell')=\int_0^{\infty}P_{n\ell}(r)rP_{n'\ell'}(r)dr
\end{equation}
is the radial integral. For electric-quadrupole lines, the $J-$file sum rule reads
\begin{eqnarray}
& &S_{E2}\left[\left(\ell^N\ell'^{N'+1}\right)\alpha J-\ell^{N+1}\ell'^{N'}\right]\nonumber\\
& &\;\;\;\;\;\;\;\;\;\;\;\;=(2J+1)\left[\frac{(N'+1)}{2\ell'+1}\langle\ell\vert\vert \mathscr{C}^{(2)}\vert\vert\ell'\rangle^2+\mathscr{C}(G^2;\alpha J)\right]J^2(n\ell,n'\ell'),\nonumber\\
& &
\end{eqnarray}
where
\begin{equation}
J(n\ell,n'\ell')=\int_0^{\infty}P_{n\ell}(r)r^2P_{n'\ell'}(r)dr
\end{equation}
and
\begin{equation}
\langle\ell\vert\vert \mathscr{C}^{(2)}\vert\vert\ell'\rangle^2=\frac{3\ell_>(\ell_>-1)}{2(2\ell_>-1)}
\end{equation}
if $\vert\ell-\ell'\vert=2$ and
\begin{equation}
\langle\ell\vert\vert \mathscr{C}^{(2)}\vert\vert\ell\rangle^2=\frac{\ell(\ell+1)(2\ell+1)}{(2\ell-1)(2\ell+3)}.
\end{equation}
$\mathscr{C}(G^2;\alpha J)$ is the coefficient of the $G^2$ Slater integral in the energy of the $\alpha J$ level. The $G^1$ integrals plays also a major role in the characterization of the singlet-triplet mixing in configuration-interaction studies \cite{Curtis1989,Curtis2000, Pain2017}.
Very few things are known about the general properties of exchange Slater integrals \cite{Bacher1933,Condon1935}. Since the integrand of the right-hand side of Eq. (\ref{fk}) is positive for every $(r_1,r_2)$ and gets smaller as $k$ increases, one has
\begin{equation}
F^0 > F^1 > F^2 > \cdots > 0.
\end{equation}
It was also proven by Racah \cite{Racah1942}, that $G^k$ is always positive and that 
\begin{equation}
\frac{G^k}{2k+1} > \frac{G^{k+1}}{2k+3}>0. 
\end{equation}
However, as stated by Cowan \cite{Cowan1981}, for ``realistic forms'' of the radial functions $P_i(r)$, the exchange integrals usually (although not always) satisfy
\begin{equation}
G^0>G^1>\cdots>0.
\end{equation} 
In the present work, we push further the integral approach used by Racah in his seminal paper, to derive new identities for exchange Slater integrals. We derive, following the procedure introduced by Racah, two inequalities, involving respectively three and four consecutive (with respect to their order $k$) exchange Slater integrals. Such relations can prove useful to detect regularities in complex atomic spectra \cite{Pain2013} and classify energy levels. They can bring an insight into the relative importance of the contribution to the energy of a configuration, and provide information about the relative importance of emissive zones for instance.

Using $G^1$ integrals as free fitting parameters and taking advantage of their relative importance for different configurations, Sugar interpreted the resonances in the photo-absorption spectra of lanthanides near the $4d$ absorption edge as transitions $4d^{10}4f^N-4d^94f^{N+1}$ \cite{Sugar1972}. Applying the Racah tensor operator methods \cite{Racah1943,Racah1949}, the author evaluated the angular part of the energy matrices and left the radial integrals as parameters to be adjusted to fit the experimental data. For that purpose, Sugar resorted to a simplification proposed by Fano, Prats and Goldschmidt for reducing the recoupling, required to associate the interacting electrons, to one recoupling coefficient \cite{Fano1963}. This coefficient was then obtained by the diagrammatic methods of Yutsis \emph{et al.} \cite{Yutsis1962,Massot1966} (due to a double transliteration, into Cyrillic and back into roman, the original name Jucys is often replaced, in the literature, by its more nearly phonetic version, Yutsis \cite{Fano1974}). Sugar's calculations of the $4d^94f^{N+1}$ energy levels and relative $gf$ (oscillator strength multiplied by the degeneracy of the initial level) values for these transitions were found to be in good agreement with the measured absorption spectra.

It is instructive to study how the energies of the levels change when some exchange Slater integrals dominates, like in the vicinity of the so-called PH (particle-hole) coupling, where only the exchange Slater integrals $G^k$ are non zero (in the case where $G^1$ highly dominates, it is referred to as the $G^1$ coupling). For the configurations with one vacancy $n\ell^{4\ell+1}n(\ell+1)^{N_2}$ it was suggested \cite{Sugar1972} (and generalized for any configuration with two open shells $n\ell^{N_1} n(\ell+1)^{N_2}$ \cite{Karazija1985}) to classify the energy levels in a way which preserves $n\ell^{4\ell+1}n(\ell+1)$ parentage. Such a basis can be obtained by diagonalizing the matrix of the coefficient of the main Coulomb exchange integral $G^1$ \cite{Bernotas2001}. 

The new inequalities presented here can not be obtained as simple combination of the existing ones. Racah's results are briefly recalled in section \ref{sec2}, and the two new relations in section \ref{sec3}, together with the technique we have adopted, which can be generalized to higher numbers of exchange Slater integrals. Possible extensions to the relativistic case are discussed in section \ref{sec4}.

\section{Racah's inequalities}\label{sec2}

In the following we assume that the wavefunction and its derivatives cancel at infinity. The exchange Slater integral $G^k$ can be put in the form
\begin{equation}
G^k=\int_0^{\infty}\int_0^{\infty}\frac{r_<^k}{r_>^{k+1}}f(r_1)f(r_2)dr_1dr_2
\end{equation}
or equivalently
\begin{equation}\label{gkf}
G^k=\int_0^{\infty}f(x)\phi_k(x)dx
\end{equation}
with
\begin{equation}\label{phik}
\phi_k(x)=x^{-k-1}\int_0^{x}y^kf(y)dy+x^{k}\int_x^{\infty}y^{-k-1}f(y)dy.
\end{equation}

\subsection{Case of one exchange Slater integral $G^k$}

Using equation (\ref{phik}), one obtains (the prime symbol denotes the derivative):
\begin{equation}
f(x)=-\frac{x^{k+1}}{2k+1}\left[x^{-2k}\left(x^{k+1}\phi_k(x)\right)'\right]'
\end{equation}
and integrating by parts the right-hand side of Eq. (\ref{gkf}) leads to
\begin{equation}\label{rac1}
G^k=\frac{1}{2k+1}\int_0^{\infty}x^{-2k}\left[\left(x^{k+1}\phi_k(x)\right)'\right]^2dx\geq 0,
\end{equation}
which is Eq. (107) of Ref. \cite{Racah1942}.

\subsection{Case of two exchange Slater integrals $G^k$}

In the same way, setting
\begin{equation}
\psi_k(x)=\frac{1}{2k+1}\phi_k(x)-\frac{1}{2k+3}\phi_{k+1}(x),
\end{equation}
one has
\begin{equation}\label{cas2}
\frac{G^k}{2k+1}-\frac{G^{k+1}}{2k+3}=\int_0^{\infty}f(x)\psi_k(x)dx
\end{equation}
and
\begin{equation}
f(x)=\frac{x^{k+2}}{(2k+2)}\left[x^{-2k}\left(x^ {k+2}\psi_k(x)\right)''\right]''.
\end{equation}
Since one has
\begin{eqnarray}
\left(x^{k+2}\psi(x)\right)'&=&\alpha_k\left[\int_0^xy^kf(y)dy.+(2k+2)x^{2k+1}\int_x^{\infty}y^{-k-1}f(y)dy\right]\nonumber\\
& &+\beta_k(2k+1)x^{2k+2}\int_x^{\infty}y^{-k-2}f(y)dy,
\end{eqnarray}
as well as
\begin{eqnarray}
\left(x^{k+2}\psi_k(x)\right)''&=&\alpha_k\left[x^{-k}f(x)+(2k+2)(2k+1)x^{2k}\int_x^{\infty}y^{-k-1}f(y)dy\right.\nonumber\\
& &\left.-(2k+2)x^kf(x)\right]\nonumber\\
& &+\beta_k\left[(2k+3)(2k+2)x^{2k+1}\int_x^{\infty}y^{-k-2}f(y)dy\right.\nonumber\\
& &\left.-(2k+3)x^kf(x)\right],
\end{eqnarray}
one gets
\begin{eqnarray}
\left[x^{-2k}\left(x^ {k+2}\psi_k(x)\right)''\right]'&=&\alpha_k\left[-kx^{-k-1}f(x)+x^{-k}f'(x)\right.\nonumber\\
& &\left.-(2k+2)(2k+1)x^{-k-1}f(x)\right.\nonumber\\
& &\left.+k(2k+2)x^{-k-1}f(x)-(2k+2)x^{-k}f'(x)\right]\nonumber\\
& &+\beta_k\left[(2k+3)(2k+2)\int_x^{\infty}y^{-k-2}f(y)dy\right.\nonumber\\
& &-(2k+3)(2k+2)x^{-k-1}f(x)\nonumber\\
& &\left.+k(2k+3)x^{-k-1}f(x)-(2k+3)x^{-k}f'(x)\right],
\end{eqnarray}
yielding, after a double integration by parts of the right-hand side of Eq. (\ref{cas2}):
\begin{equation}
\frac{G^k}{2k+1}-\frac{G^{k+1}}{2k+3}=\frac{1}{(2k+2)}\int_0^{\infty}x^{-2k}\left[\left(x^{k+2}\psi_k(x)\right)''\right]^2dx
\end{equation}
\emph{i.e.}
\begin{equation}\label{rac2}
\frac{G^k}{2k+1}-\frac{G^{k+1}}{2k+3}\geq 0,
\end{equation}
which is exactly Eq. (110) of Ref. \cite{Racah1942}.

\section{Further inequalities}\label{sec3}

\subsection{Case of three exchange Slater integrals $G^k$}\label{subsec31}

Taking
\begin{eqnarray}
\chi_k(x)&=&\frac{1}{4(2k+1)(k+1)}\phi_k(x)-\frac{1}{4(k+2)(k+1)}\phi_{k+1}(x)\nonumber\\
& &+\frac{1}{4(k+2)(2k+5)}\phi_{k+2}(x),
\end{eqnarray}
one gets
\begin{equation}
f(x)=-\frac{x^{k+3}}{2k+3}\left[x^{-2k}\left(x^{k+3}\chi_k(x)\right)^{(3)}\right]^{(3)},
\end{equation}
where superscript ${(n)}$ means the $n^{th}$ derivative. It is easy to show that
\begin{eqnarray}
\left(x^{k+3}\psi(x)\right)'&=&\alpha_k\left[2x\int_0^xy^{k}f(y)dy+(2k+3)x^{2k+2}\int_x^{\infty}y^{-k-1}f(y)dy\right]\nonumber\\
& &+\beta_k\left[\int_0^xy^{k+1}f(y)dy+(2k+4)x^{2k+3}\int_x^{\infty}y^{-k-2}f(y)dy\right]\nonumber\\
& &+\gamma_k(2k+5)x^{2k+4}\int_x^{\infty}y^{-k-3}f(y)dy,
\end{eqnarray}
leading to
\begin{eqnarray}
\left(x^{k+3}\psi(x)\right)''&=&\alpha_k\left[2\int_0^xy^{k}f(y)dy\right.\nonumber\\
& &\left.+2x^{k+1}f(x)+(2k+3)(2k+2)x^{2k+1}\int_x^{\infty}y^{-k-1}f(y)dy\right.\nonumber\\
& &\left.-(2k+3)x^{k+1}f(x)\right]\nonumber\\
& &+\beta_k\left[x^{k+1}f(x)+(2k+4)(2k+3)x^{2k+2}\int_x^{\infty}y^{-k-2}f(y)dy\right.\nonumber\\
& &\left.-(2k+4)x^{k+1}f(x)\right]\nonumber\\
& &+\gamma_k\left[(2k+5)(2k+4)x^{2k+3}\int_x^{\infty}y^{-k-3}f(y)dy\right.\nonumber\\
& &\left.-(2k+5)x^{k+1}f(x)\right],
\end{eqnarray}
as well as
\begin{eqnarray}
\left(x^{k+3}\psi(x)\right)^{(3)}&=&\alpha_k\left[2x^kf(x)+2(k+1)x^kf(x)+2x^{k+1}f'(x)\right.\nonumber\\
& &\left.+(2k+3)(2k+2)(2k+1)x^{2k}\int_x^{\infty}y^{-k-1}f(y)dy\right.\nonumber\\
& &\left.-(2k+3)(2k+2)x^kf(x)-(2k+3)(k+1)x^kf(x)\right.\nonumber\\
& &\left.-(2k+3)x^{k+1}f'(x)\right]\nonumber\\
& &+\beta_k\left[(k+1)x^{k}f(x)+x^{k+1}f'(x)\right.\nonumber\\
& &\left.+(2k+4)(2k+3)(2k+2)x^{2k+1}\int_x^{\infty}y^{-k-2}f(y)dy\right.\nonumber\\
& &\left.-(2k+4)(2k+3)(k+1)x^{k}f(x)\right.\nonumber\\
& &\left.-(2k+4)(k+1)x^kf(x)-(2k+4)x^{k+1}f'(x)\right]\nonumber\\
& &+\gamma_k\left[(2k+5)(2k+4)(2k+3)x^{2k+2}\int_x^{\infty}y^{-k-3}f(y)dy\right.\nonumber\\
& &\left.-(2k+5)(2k+4)x^{k}f(x)\right.\nonumber\\
& &\left.-(2k+5)(k+1)x^kf(x)-(2k+5)x^{k+1}f'(x)\right].\nonumber\\
& &
\end{eqnarray}

After a triple integration by parts of the right-hand side of
\begin{eqnarray}\label{cas3}
& &\frac{1}{2}\frac{G^k}{(2k+1)(2k+2)}-\frac{G^{k+1}}{(2k+2)(2k+4)}+\frac{1}{2}\frac{G^{k+2}}{(2k+4)(2k+5)}\nonumber\\
&=&\;\;\;\;-\int_0^{\infty}f(x)\chi_k(x)dx,
\end{eqnarray}
one obtains
\begin{eqnarray}\label{coeff3}
& &\frac{1}{2}\frac{G^k}{(2k+1)(2k+2)}-\frac{G^{k+1}}{(2k+2)(2k+4)}+\frac{1}{2}\frac{G^{k+2}}{(2k+4)(2k+5)}\nonumber\\
&=&\frac{1}{(2k+3)}\int_0^{\infty}x^{-2k}\left[\left(x^{k+3}\chi_k(x)\right)^{(3)}\right]^2dx
\end{eqnarray}
implying
\begin{equation}
\frac{G^k}{(2k+1)(2k+2)}-2\frac{G^{k+1}}{(2k+2)(2k+4)}+\frac{G^{k+2}}{(2k+4)(2k+5)}\geq 0.
\end{equation}
Such a relation, which is the first main result of the present article, can not be deduced from the inequalities (\ref{rac1}) and (\ref{rac2}) published by Racah.

\subsection{Case of four exchange Slater integrals $G^k$}\label{subsec32}

In a similar manner, setting
\begin{eqnarray}
\eta_k(x)&=&\frac{1}{6(2k+1)(2k+2)(2k+3)}\phi_k(x)\nonumber\\
& &-\frac{1}{2(2k+2)(2k+3)(2k+5)}\phi_{k+1}(x)\nonumber\\
& &+\frac{1}{2(2k+3)(2k+5)(2k+6)}\phi_{k+2}(x)\nonumber\\
& &-\frac{1}{6(2k+5)(2k+6)(2k+7)}\phi_{k+3}(x),
\end{eqnarray}
one has, {\it mutatis mutandis}
\begin{equation}
f(x)=\frac{x^{k+4}}{2k+4}\left[x^{-2k}\left(x^{k+4}\eta_k(x)\right)^{(4)}\right]^{(4)}
\end{equation}
and since
\begin{eqnarray}\label{cas4}
& &\frac{G^k}{6(2k+1)(2k+2)(2k+3)}-\frac{G^{k+1}}{2(2k+2)(2k+3)(2k+5)}\nonumber\\
& &+\frac{G^{k+2}}{2(2k+3)(2k+5)(2k+6)}-\frac{G^{k+3}}{6(2k+5)(2k+6)(2k+7)}\nonumber\\
&=&\int_0^{\infty}f(x)\eta_k(x)dx,
\end{eqnarray}
one obtains, after four successive integrations by parts of the integral in the right-hand-side of the preceding equation
\begin{eqnarray}\label{coeff4}
& &\frac{G^k}{6(2k+1)(2k+2)(2k+3)}-\frac{G^{k+1}}{2(2k+2)(2k+3)(2k+5)}\nonumber\\
& &+\frac{G^{k+2}}{2(2k+3)(2k+5)(2k+6)}-\frac{G^{k+3}}{6(2k+5)(2k+6)(2k+7)}\nonumber\\
&=&\frac{1}{2k+4}\int_0^{\infty}x^{-2k}\left[\left(x^{k+4}\eta_k(x)\right)^{(4)}\right]^2dx,
\end{eqnarray}
leading, the quantity in the integral being positive over the whole space, to
\begin{eqnarray}
& &\frac{G^k}{(2k+1)(2k+2)(2k+3)}-3\frac{G^{k+1}}{(2k+2)(2k+3)(2k+5)}\nonumber\\
& &\;\;\;\;+3\frac{G^{k+2}}{(2k+3)(2k+5)(2k+6)}-\frac{G^{k+3}}{(2k+5)(2k+6)(2k+7)}\geq 0,
\end{eqnarray}
which constitutes the second main result of the present work.

\section{Method and generalization}\label{sec4}

Although we could not find any generalization of such formulas for an arbitrary number of exchange Slater integrals, the procedure used for the two latter cases (three and four integrals respectively) can be applied for higher numbers of Slater integrals, using a computer algebra system such as Mathematica \cite{Mathematica}.

The most difficult point is to find the coefficients in front of the $G^k$ Slater integrals. For instance, in the three-integral case, we set
\begin{equation}
\chi_k(x)=\alpha_k\phi_k(x)+\beta_k\phi_{k+1}(x)+\gamma_k\phi_{k+2}(x),
\end{equation}
then calculate
\begin{eqnarray}
-\frac{x^{k+3}}{2k+3}\left[x^{-2k}\left(x^{k+3}\chi_k(x)\right)^{(3)}\right]^{(3)}&=&A_kf(x)+B_kf'(x)+C_kf''(x)\nonumber\\
& &+D_kf^{(3)}(x)+E_kf^{(4)}(x)
\end{eqnarray}
and solve the following system:
\begin{equation}
\left\{\begin{array}{l}
A_k=1\\
B_k=C_k=D_k=E_k=0,
\end{array}
\right.
\end{equation}
which is redundant and boils down to the set of three independent equations
\begin{equation}
\left\{\begin{array}{l}
A_k=-\left[-(k+2)(k+3)(2k+1)\alpha_k+k(k+3)(2k+3)\beta_k\right.\\
\;\;\;\;\;\;\;\;\left.+k(k+1)(2k+5)\gamma_k\right]/(2k+3)=1,\\
B_k=(k+1)(k+3)(2k+1)\alpha_k+(k+1)(k+2)(2k+3)\beta_k\\
\;\;\;\;\;\;\;\;+k(k+2)(2k+5)\gamma_k=0,\\
D_k=(2k+1)\alpha_k+(2k+3)\beta_k+(2k+5)\gamma_k=0,
\end{array}
\right.
\end{equation}
yielding
\begin{equation}
\left\{\begin{array}{l}
\alpha_k=\displaystyle\frac{1}{4(k+1)(2k+1)}\\
\beta_k=-\displaystyle\frac{1}{4(k+2)(k+1)}\\
\gamma_k=\displaystyle\frac{1}{4(k+2)(2k+5)}
\end{array}
\right.
\end{equation}
as indicated in Eq. (\ref{coeff3}).

Following the same procedure, in the four-integral case, we set
\begin{equation}
\eta_k(x)=\alpha_k\phi_k(x)+\beta_k\phi_{k+1}(x)+\gamma_k\phi_{k+2}(x)+\delta_k\phi_{k+3}(x)
\end{equation}
and calculate
\begin{eqnarray}
\frac{x^{k+4}}{2k+4}\left[x^{-2k}\left(x^{k+4}\chi_k(x)\right)^{(4)}\right]^{(4)}&=&A_kf(x)+B_kf'(x)+C_kf''(x)\nonumber\\
& &+D_kf^{(3)}(x)+E_kf^{(4)}(x)\nonumber\\
& &+H_kf^{(5)}(x)+I_kf^{(6)}(x).
\end{eqnarray}
Here also, one has to solve a system 
\begin{equation}
\left\{\begin{array}{l}
A_k=1\\
B_k=C_k=D_k=E_k=H_k=I_k=0,
\end{array}\right.
\end{equation}
which is redundant and reduces to the set of four independent equations
\begin{equation}
\left\{\begin{array}{l}
A_k=\left[(k+1)(k+2)(k+3)^2(k+4)(2k+1)/2\right]\alpha_k\\
\;\;\;\;\;\;\;\;+\left[k(k+1)(k+3)^2(k+4)(2k+3)/2\right]\beta_k\\
\;\;\;\;\;\;\;\;+\left[k(k+1)^2(k+3)(k+4)(2k+5)/2\right]\gamma_k\\
\;\;\;\;\;\;\;\;+\left[k(k+1)^2(k+2)(k+3)(2k+7)/2\right]\delta_k=1,\\
B_k=\left[3(k+1)(k+3)(k+4)(2k+1)\right]\alpha_k\\
\;\;\;\;\;\;\;\;+\left[(k+1)(k+4)(2k+3)(3k+5)\right]\beta_k\\
\;\;\;\;\;\;\;\;+\left[k(k+3)(2k+5)(3k+7)\right]\gamma_k\\
\;\;\;\;\;\;\;\;+\left[3k(k+1)(k+3)(2k+7)\right]\delta_k=0,\\
D_k=3(2k+1)\left[k(k+5)-1\right]\alpha_k\\
\;\;\;\;\;\;\;\;+(2k+3)\left[-5+k(3k+13)\right]\beta_k\\
\;\;\;\;\;\;\;\;+(2k+5)\left[-9+k(3k+11)\right]\gamma_k\\
\;\;\;\;\;\;\;\;+3(2k+7)\left[-5+k(k+3)\right]\delta_k=0,\\
H_k=(2k+1)\alpha_k+(2k+3)\beta_k+(2k+5)\gamma_k+(2k+7)\delta_k=0,
\end{array}
\right.
\end{equation}
yielding
\begin{equation}
\left\{\begin{array}{l}
\alpha_k=\displaystyle\frac{1}{12(k+1)(2k+1)(2k+3)}\\
\beta_k=-\displaystyle\frac{1}{4(k+1)(2k+3)(2k+5)}\\
\gamma_k=\displaystyle\frac{1}{4(k+3)(2k+3)(2k+5)}\\
\delta_k=-\displaystyle\frac{1}{12(k+3)(2k+5)(2k+7)}
\end{array}
\right.
\end{equation}
and subsequently Eq. (\ref{coeff4}).

\section{Relativistic effects}

In many spectroscopic problems it is interesting to deduce non-relativistic ($n\ell$ representation) radial integrals from the Breit-Dirac equations of a relativistic ($n\ell j$ representation) central-field or Hartree-Fock calculation. Such a derivation was proposed by Larkins \cite{Larkins1976}, who published some expansions of the non-relativistic $F^k$ and $G^k$ Slater integrals in terms of their relativistic counterparts. To obtain the $F^k(n\ell,n\ell)$ and $G^k(n\ell,n'\ell')$ expansions, he equated the average energies of any given electronic configuration in both the non-relativistic and relativistic schemes. This was not possible for the $F^k(n\ell,n'\ell')$ integrals (with $k \ne 0$), which do not occur in the average energies. Bauche \emph{et al.} proposed new expansions for these $F^k$ integrals, sharing some interesting properties with Larkins’s expansions for the $G^k$ integrals \cite{Bauche1982}. Their method, based on the equivalent-operator formalism \cite{Armstrong1966,Armstrong1968}, involves a linear approximation for obtaining the Slater integrals; as concerns the exchange integrals, their results agree with those previously published by Larkins, but some differences appear for the direct integrals. The linear approximation is very well adapted to the study of moderately ionized atoms (typically for ionization degrees smaller than 40), for which it is sufficient to keep the contributions of order $\alpha^2$ ($\alpha$ being the fine-structure constant) in the non-relativistic limit, and for which the spin-orbit interaction largely predominates over the other magnetic ones. If the relativistic Slater integrals do not fulfil the linear approximation, the complete expressions provided in Refs. \cite{Armstrong1966,Armstrong1968} must be used. The non-relativistic exchange Slater integrals $G_{\mathrm{NR}}^k(\ell,\ell')$ can be expressed in terms of the relativistic ones $G_{\mathrm{R}}^k(j,j')$ by
\begin{equation}\label{fiftyfour}
G_{\mathrm{NR}}^k(\ell,\ell')=\frac{\sum_{jj'}w_{jj'}G_{\mathrm{R}}^k(jj')}{\sum_{jj'}w_{jj'}},
\end{equation}
where $w_{jj'}=(2j+1)(2j'+1)$ is the statistical weight. For instance, one has, for $d$ ($\ell=2$) and $f$ ($\ell=3$) orbitals:
\begin{equation}
G_{\mathrm{NR}}^1(d,f)=\frac{1}{35}\left[6~G^1\left(\frac{3}{2},\frac{5}{2}\right)+12~G^1\left(\frac{5}{2},\frac{7}{2}\right)+8~G^1\left(\frac{3}{2},\frac{7}{2}\right)+9~G^1\left(\frac{5}{2},\frac{5}{2}\right)\right].
\end{equation}

In Ref. \cite{Bauche1982}, the spin-orbit integrals have been defined by averaging the two-particle relativistic operators (Coulomb and Breit interactions) over all the states of a given configuration. It is important to stress that the contributions coming from the Breit interaction cannot be neglected; already for moderately ionized atoms (typically ionization larger than 10), they are approximately twice the contributions from the exchange part of the Coulomb interaction.

Using the abovementioned correspondence (see Eq. (\ref{fiftyfour})), our new identities for exchange Slater integrals can be applied directly.

However, the use of relativistic wave functions becomes necessary not only at high values of effective nuclear charge, but also for the outer shells of heavy atoms due to a shrinking of inner shells \cite{Jonauskas2003}.

Using the equivalent relativistic operator and the correspondence of its terms to the operators in the Breit–Pauli approximation, the relativistic analogs for the integrals of Coulomb, spin–contact, spin–orbit, spin–spin interactions were obtained by Jonauskas and Karazija \cite{Jonauskas2003}. In spherical polar coordinates, the solutions can be written in the form
\begin{equation}
\phi_{n\kappa m}(r) = \frac{1}{r}\left(\begin{array}{c}
P_{n\kappa j}(r)\chi_{\kappa, m}\\
iQ_{n\kappa j}(r)\chi_{-\kappa, m}
\end{array}
\right), 
\end{equation}
where $\chi_{\kappa, m}$ is an angular spin function \cite{Aoyagi1977,Grant2007}, and $P$ and $Q$ are the large and small components, respectively. The total angular momentum $j$ is given by $j=\ell-\nu/2$, where $\nu=\pm 1$. The quantum number $\kappa$ is given by $\kappa=(j+1/2)\nu$. The radial Slater integrals are defined by $F^k=R^k(a,b,a,b)$ and $G^k=R^k(a,b,b,a)$, where
\begin{eqnarray}
R^k(a,b,c,d)&=&\int_0^{\infty}\int_0^{\infty}\left[P_a(r_1)P_c(r_1)+Q_a(r_1)Q_c(r_1)\right]\nonumber\\
& &\times\frac{r_<^k}{r_>^{k+1}}\left[P_b(r_2)P_d(r_2)+Q_b(r_2)Q_d(r_2)\right]dr_1dr_2,
\end{eqnarray}
with $r_< = \min(r_1, r_2)$ and $r_> = \max(r_1, r_2)$.

Therefore, it should be possible, {\it albeit} at the cost of more complex calculations, to derive inequalities similar to the ones of the present work, following the same procedure. 

\section{Conclusion}\label{sec6}

We found new inequalities satisfied by exchange Slater integrals. The variations of exchange Slater integrals with respect to their order $k$ are not well known. While direct Slater integrals $F^k$ are positive and decreasing when the order increases, this is not always the case for exchange integrals $G^k$. In this work, we showed that the technique used by Racah a long time ago can be generalized, although the complexity increases, to derive further relations, and give two of them, consisting in inequalities involving respectively three and four exchange integrals. It is hoped that such relations will be useful to check the validity of advanced computations and to shed light on interesting features of atomic spectra, such as propensity rules, which are not only of fundamental nature, but can also give rise to the development of statistical models. In the future, we plan to try to derive similar inequalities for the relativistic Slater integrals, involving Dirac wavefunctions. Moreover, independently of the relativistic character, it might be interesting to directly deal with the general $R^k$ integrals, sometimes referred to as ``configuration-interaction integrals'' or ``four-parameter'' radial integrals.


\section*{References}

\begin{thebibliography}{99}

\bibitem{Sobelman1992} Sobelman, I.I.: Atomic spectra and radiative transitions. Springer, Berlin, Heidelberg (1992).

\bibitem{Cowan1981} Cowan, R.D.: The theory of atomic structure and spectra. University of California Press, Berkeley (1981).

\bibitem{Gaunt1929} Gaunt, J. A.: On the triplets of helium, Philos. Trans. Roy. Soc. (London) Ser. A {\bf 228},
151-196 (1929).

\bibitem{Xu1996} Yu-Lin Xu, Fast evaluation of the Gaunt coefficients, Math. Comput. {\bf 65}, 1601–1612 (1996).

% Reference for Slater integrals:
\bibitem{Slater1929} Slater, J.C.: The theory of complex spectra. Phys. Rev. {\bf 34}, 1293-1322 (1929).

\bibitem{Naqvi1964} Naqvi, A.M.: Calculations and applications of screened hydrogenic wavefunctions. J. Quant. Spectrosc. Radiat. Transfer. {\bf 4}, 597-615 (1964).

\bibitem{Ruano2013} Ruano, F.H., Rubiano, J.G., Mendoza, M.A., Gil, J.M., Rodr\'iguez, R., Florido, R., Martel, P., M\'inguez, E.: Relativistic screened hydrogenic radial integrals. J. Quant. Spectrosc. Radiat. Transfer {\bf 117}, 123-132 (2013). 

\bibitem{Hey2017} Hey, J.D.: On forms of the Coulomb approximation as a useful source of atomic data for the spectroscopy of astrophysical and fusion plasmas. J. Phys. B: At. Mol. Opt. Phys. {\bf 50}, 065701 (2017).

\bibitem{Condon1935} Condon, E.U., Shortley, G.H.: The Theory of Atomic Spectra. Cambridge University Press, New York and London (1935).

\bibitem{OSullivan1999} O'Sullivan, G., Carroll, P.K., Dunne, P., Faulkner, R., McGuinness, C., Murphy, N.: Supercomplex spectra and continuum emission from rare-earth ions: Sm, a case study. J. Phys. B: At., Mol. Opt. Phys. {\bf 32}, 1893-1922 (1999).

\bibitem{Bauche-Arnoult2000} Bauche-Arnoult, C., Bauche, J., Wyart, J.-F., Fournier, K.B.: Effects of the exchange Slater integrals on the shapes of transition arrays. J. Quant. Spectrosc. Radiat. Transfer {\bf 65}, 57-70 (2000).

\bibitem{Bauche-Arnoult1983} Bauche, J., Bauche-Arnoult, C. Luc-Koenig, E., Wyart, J.-F., Klapisch, M.: Emissive zones of complex atomic configurations in highly ionized atoms. Phys. Rev. A {\bf 28}, 829-835 (1983).

\bibitem{Bauche2015} Bauche, J., Bauche-Arnoult, C., Peyrusse, O.: Atomic properties in hot plasmas: From levels to superconfigurations. Springer (2015).

\bibitem{Curtis1989} Curtis, L.J.: Semiempirical specification of singlet-triplet mixing angles, oscillator strengths, and $g$ factors in $nsn'l$, $nsn'p^5$, $np^2$, and $np^4$ configurations. Phys. Rev. A {\bf 40}, 6958-6968 (1989).

\bibitem{Curtis2000} Curtis, L.J.: Branching fractions and transition probabilities for Ga II, In II and Tl II from measured lifetime and energy level data. Phys. Scr. {\bf 62}, 31-35 (2000).

\bibitem{Pain2017} Pain, J.-C., Gilleron, F.: Statistical properties of levels and lines in complex spectra: A tribute to Jacques Bauche and Claire Bauche-Arnoult. AIP Conf. Proc. {\bf 1811}, 050003 (2017).

\bibitem{Bacher1933} Bacher, R.F.: The interaction of configurations: $sd - p^2$. Phys. Rev. {\bf 43}, 264-269 (1933).

\bibitem{Racah1942} Racah, G.: Theory of complex spectra. II. Phys. Rev. {\bf 62}, 438-462 (1942).

\bibitem{Pain2013} Pain, J.-C.: Regularities and symmetries in atomic structure and spectra, High Energy Density Phys. {\bf 9}, 392-401 (2013).

\bibitem{Sugar1972} Sugar, J.: Potential-barrier effects in photoabsorption. II. Interpretation of photoabsorption resonances in lanthanide metals at the $4d$-electron threshold. Phys Rev. B {\bf 5}, 1785-1792 (1972).

\bibitem{Racah1943} Racah, G.: Theory of complex spectra. III. Phys. Rev. {\bf 63}, 367-382 (1943).

\bibitem{Racah1949} Racah, G.: Theory of complex spectra. IV. Phys. Rev. {\bf 76}, 1352-1365 (1949).

\bibitem{Fano1963} Fano, U., Prats, F.,Goldschmidt, Z.: Interaction Matrix Element in a Shell Model. Phys. Rev. {\bf 129}, 2643-2652 (1963).

\bibitem{Yutsis1962} Jucys, A.P., Levinsonas, J.B., Vanagas, V.V.: {\it Mathematical Apparatus of the Theory of Angular Momentum}, NASA technical translation, Israel Program for Scientific Translations (1962).

\bibitem{Massot1966} Massot, J.N., El-Baz E., Lafoucriere, J.: M\'ethode graphique de sommation et d'int\'egration des harmoniques sph\'eriques. Nucl. Phys. {\bf 83}, 449-459 (1966).

\bibitem{Fano1974} Fano, U.: Adolfas Jucys. Phys. Today {\bf 27}, 72 (1974).

\bibitem{Karazija1985} Karazija, R. Liet. Fiz. Rink. {\bf 25}, 32 Engl. transl. Karazija, R., Ku$\mathrm{\check{c}}$as, S Sov. Phys. Collection {\bf 25}, 23 (1985).

\bibitem{Bernotas2001} Bernotas, A., Karazija, R.: Additional selection rule for some emission, photoexcitation and Auger spectra. J. Phys. B: At. Mol. Opt. Phys. {\bf 34}, L741-L747 (2001).

\bibitem{Mathematica} Wolfram Research, Inc., Mathematica, Version 13.2, Champaign, IL (2022).

\bibitem{Larkins1976} Larkins, F.P.: Relativistic LS multiplet energies for atoms and ions. J. Phys. B: At. Mol. Phys. {\bf 9}, 37-46 (1976).

\bibitem{Bauche1982} Bauche, J., Bauche-Arnoult, C., Luc-Koenig, E., Klapisch, M.: Non-relativistic energies from relativistic radial integrals in atoms and ions. J. Phys. B: At. Mol. Phys. {\bf 15}, 2325-2338 (1982).

\bibitem{Armstrong1966} Armstrong Jr, L.: Relativistic effects in atomic fine structure. J. Math. Phys. {\bf 7}, 1891-1899 (1966).

\bibitem{Armstrong1968} Armstrong Jr, L.: Relativistic effects in atomic fine structure. II. J. Math. Phys. {\bf 9}, 1083-1086 (1968).

\bibitem{Jonauskas2003} Jonauskas, V., Karazija, R.: General relations between radial integrals in nonrelativistic and relativistic calculation schemes. J. Math. Phys. {\bf 44}, 1660-1665 (2003).

\bibitem{Aoyagi1977} Aoyagi, M., Chen, M.H., Crasemann, B., Huang, K.-N., Mark, H.: Relativistic electrostatic Slater integrals, $2 \leq Z \leq 106$, Ames Research Center, NASA Technical Report R-464 (1977).

\bibitem{Grant2007} Grant, I.P.: Relativistic quantum theory of atoms and molecules. Springer, New York (2007).

\end{thebibliography}
\end{document}